\documentclass[cits]{PoS}
\usepackage{graphicx}
\usepackage{txfonts}

\def\fs{\hbox{$.\!\!^{\rm s}$}}

\def\farcm{\hbox{$.\mkern-4mu^\prime$}}
\def\farcs{\hbox{$.\!\!^{\prime\prime}$}}

\title{Multi-frequency radio observations of BAL quasar 1045+352}

\ShortTitle{Multi-frequency radio observations of BAL quasar 1045+352}

\author{\speaker{Magdalena Kunert-Bajraszewska}\\
        Toru\'n Centre for Astronomy, N. Copernicus University, 87-100
Toru\'n, Poland\\
        E-mail: \email{magda@astro.uni.torun.pl}}

\author{Andrzej Marecki\\
        Toru\'n Centre for Astronomy, N. Copernicus University, 87-100
Toru\'n, Poland\\
        E-mail: \email{amr@astro.uni.torun.pl}}

\abstract{Multi-frequency 1.7, 5 and 8.4-GHz VLBA observations of a
radio-loud broad absorption line (BAL) quasar 1045+352 are presented. 
It is a young compact steep spectrum (CSS) object and its
asymmetric, two-sided morphology on a scale of several hundred parsecs,
extending in two different directions, may suggest intermittent activity.
The young age and unusual morphology of 1045+352 are arguments in favour
of an evolution scenario for BAL quasars, in which the BAL features appear
at a very early stage of their evolution. 
}

\FullConference{8th European VLBI Network Symposium\\
                 September 26-29, 2006\\
                 Toru\'n, Poland}

\begin{document}

\section{Introduction}

Approximately 10\% of optically selected radio-quiet
quasars display broad absorption lines (BALs) in the blue wings of the 
high ionization resonant lines
(C\,IV 1549\AA~-- high-ionization BAL (HiBAL) quasars), caused by the
outflow of gas with velocities up to $\sim0.2\,c$ \cite{hewett03}.
10\% of them also show absorption troughs in low ionization lines
(Mg\,II 2800\AA~-- low-ionization BAL (LoBAL) quasars).
Evidence has accumulated from optically selected BAL quasars, in favour of
an orientation hypothesis to explain their nature.
According to this hypothesis, BAL regions exist in both BAL and non-BAL
quasars, and that BAL quasars are normal quasars seen along a particular
line of sight, e.g. a line of sight skimming the edge of the accretion disk
or torus \cite{weymann91}. This view has been challenged by
the discovery of the existence of a large population of radio-loud BAL quasars 
\cite{broth98,becker00, menou01}.
Most of these quasars tend to be compact in the radio
domain with either a flat or steep radio spectrum. It follows
that those belonging to the latter class seem to be related to
Gigahertz-Peaked Spectrum (GPS) and Compact Steep Spectrum (CSS)
sources. This variety of
spectral indices suggests a wide range of orientations, contrary to the
interpretation favoured for optically selected quasars.

The radio morphology of BAL quasars is important because it can
serve as an inclination indicator of BALs, and therefore yields a direct
test of the orientation model. However, information about the radio
structure of BAL quasars is still very limited. Prior to 2006, only
three BAL quasars, FIRST J101614.3+520916 \cite{gregg00}, PKS 1004+13
\cite{wills99}
and LBQS 1138$-$0126 \cite{broth02} were known to have a
double-lobed FR\,II radio morphology on kiloparsec scales, although this
interpretation is doubtful for PKS 1004+13 \cite{gopal00}.
Recently, excluding PKS 1004+13, the population of FR\,II-BAL quasars has increased to ten
objects \cite{gregg06, zhou06}, although
some of them still require confirmation.
Their symmetric structures indicate an ``edge-on'' orientation, which in
turn supports an alternative hypothesis described as ``unification by time'',
with BAL quasars characterised as young or recently refuelled quasars
\cite{becker00, gregg00}. There had been only one attempt (and only at
1.6\,GHz with the EVN) to image radio
structures of the smallest (and probably the youngest) BAL quasars
\cite{jiang03}.
In this paper we present high frequency VLBA images of another very
compact BAL quasar -- 1045+352.

\section{Observations}

1045+352 belongs to the primary sample of 60 candidates for CSS sources
selected from the VLA FIRST catalogue \cite{wbhg97}. Initial observations of all the
candidates were made with MERLIN at 5\,GHz \cite{kun02} and 1.7, 5 and
8.4-GHz VLBA follow-up of 1045+352
was carried out on 13 November 2004 in a snapshot mode with
phase-referencing. The target source
scan was interleaved with a scan on a phase reference source
and the total cycle time (target and phase-reference) was
$\sim$9~minutes including telescope drive times, with $\sim$7\,minutes
actually on the target source per cycle.     

The whole data reduction process was carried out using standard AIPS
procedures. For the target source and at each frequency, the corresponding
phase-reference source was mapped and the phase errors so determined were
applied to the target sources, which were then mapped using a few cycles of
phase self-calibration and imaging. IMAGR was used to produce the final
``naturally weighted'' images shown in Fig.~\ref{1045+352_maps}.

\begin{table}[t]
\caption[]{Basic parameters of 1045+352}
\begin{tabular}{@{}l r@{}}
\hline
\hline
~~~~~Parameter & Value~~~~~\\
\hline
~~~~~Source name (B1950)     & 1045+352~~~~~\\
~~~~~Source right ascension (J2000) extracted from FIRST     & $10^{\rm
h}48^{\rm m}34\fs247$~~~~~\\
~~~~~Source declination (J2000) extracted from FIRST     & $+34^{\rm o}
57\farcm24\farcs99$~~~~~\\
~~~~~Total flux density $S_{1.4\,GHz}$~(mJy) extracted from FIRST     &
1051~~~~~\\
~~~~~log$P_{1.4\mathrm{GHz}}$~(W~${\rm Hz^{-1}}$)     & 27.65~~~~~\\
~~~~~Total flux density $S_{4.85\,GHz}$~(mJy) extracted from GB6     &
439~~~~~\\
~~~~~Spectral index $\alpha_{1.4\mathrm{GHz}}^{4.85\mathrm{GHz}}$     &
$-0.70$~~~~~\\
~~~~~Largest Angular Size measured in the 5-GHz MERLIN image     &
$0\farcs5$~~~~~\\
~~~~~Largest Linear Size ($H_0$=100${\rm\,km\,s^{-1}\,Mpc^{-1}}$,
$q_0$=0.5)~($h^{-1}$~kpc) & 2.1~~~~~\\
\hline
\end{tabular}

\label{table1}
\end{table}

\section{Results and discussion}

\begin{figure*}[t]
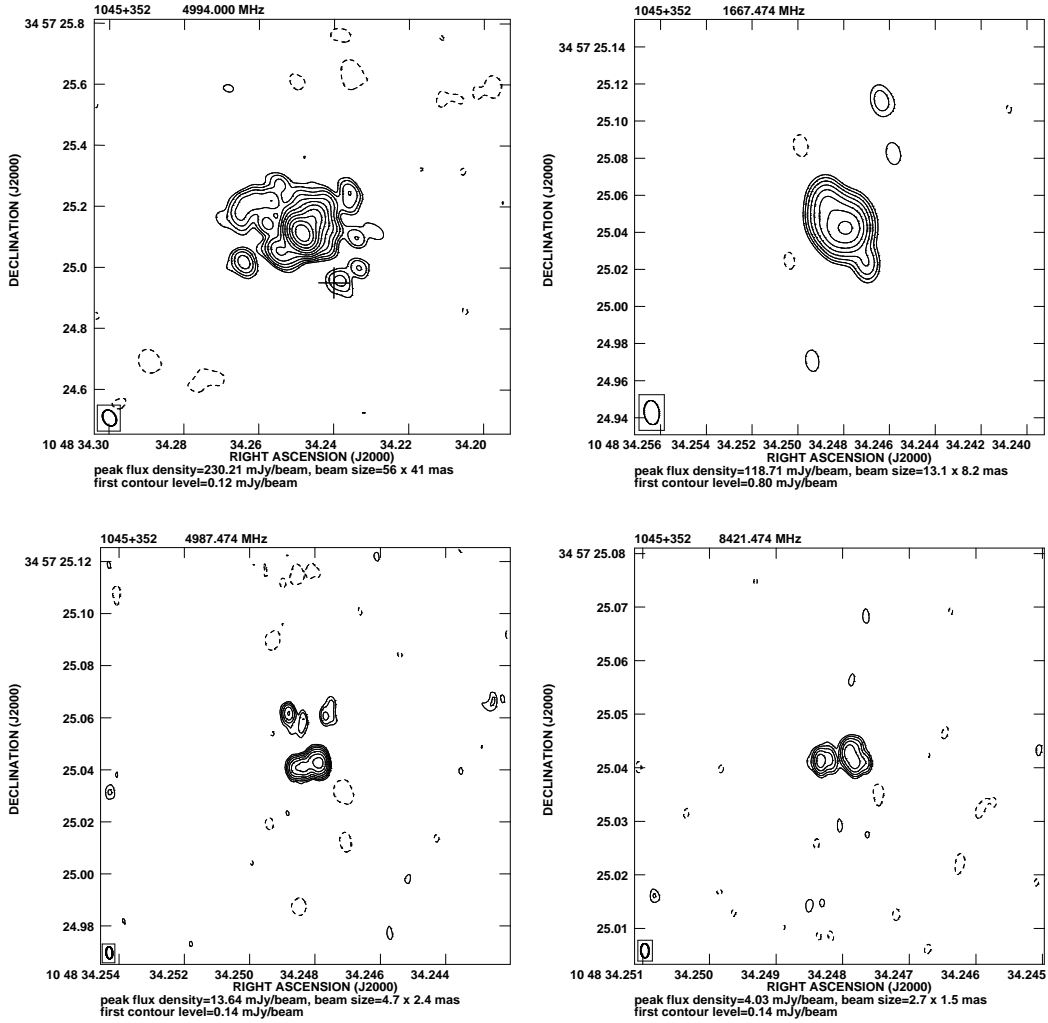

\centering
\includegraphics[width=7cm, height=7cm]{1045+352.merlin.ps}
\includegraphics[width=7cm, height=7cm]{1045+352.18cm.ps}
\includegraphics[width=7cm, height=7cm]{1045+352.6cm.ps}
\includegraphics[width=7cm, height=7cm]{1045+352.4cm.ps}
\caption{The MERLIN 5-GHz (upper left) and VLBA 1.7, 5 and 8.4-GHz maps of
1045+352. Contours increase
by a factor 2 and the first contour level corresponds to $\approx 3\sigma$.
Cross indicates the position of an optical object found using the SDSS/DR5.}
\label{1045+352_maps}
\end{figure*}

According to SDSS/DR5, 1045+352 is a galaxy at
RA=\,$10^{\rm h}48^{\rm m}34\fs242$, Dec=\,$+34^{\rm
o}57\farcm24\farcs95$,
which is marked with a cross in the MERLIN map (Fig.~\ref{1045+352_maps}). 
However, the spectral observations \cite{willott02} have shown 
1045+352 to be a quasar with a redshift
of $z=1.604$. It has also been classified as a HiBAL QSO based
upon the observed very broad
C\,IV absorption, and is a very luminous submillimetre object
with
detections at both 850\,$\mu$m and 450\,$\mu$m \cite{willott02}.

Our MERLIN and VLBA maps of 1045+352 (Fig.~\ref{1045+352_maps}) show this source to be
extended in both NE/SW and NW/SE directions. The central compact feature
with a steep spectrum
visible in all the maps is probably a radio core. The VLBA
image at 1.7\,GHz shows radio structure extended in  
a NE/SW direction, with the SW jet being weaker than
that in the NE. The NE/SW emission is visible
in the 5-GHz MERLIN image as a few compact features.
The 5-GHz VLBA image shows also a core and a one-sided jet pointing to the East.
The radio structure in the 8.4-GHz VLBA image is similar to that at
5\,GHz: an extended radio core and a jet pointing in an easterly direction.

The observed radio morphology of 1045+352 could indicate a restart
of activity accompanied by a reorientation of the jet axis with the NE/SW 
radio emission being the first phase of activity, now fading away, and
the extension in the NW/SE direction being a signature of the current
phase of activity.

A reorientation of the jet axis may result from a merger with another
black hole. It has been shown \cite{merritt02} that a rapid change of jet 
orientation can be caused by even a minor merger because of a
spin-flip of the central active black hole arising from the coalescence of
inclined binary black holes. 
Also, the Bardeen-Petterson effect \cite{bardeen} can cause a realignment of a rotating 
supermassive black hole (SMBH)
and a misaligned accretion disk \cite{liu04}, the timescale of such a realignment
being $t<10^{5}$ years. 
The interaction/realignment of a binary and its accretion disk leads to the 
development of X-shaped sources \cite{liu04}.
1045+352 is not a typical X-shaped source, however, the
realignment of a rotating SMBH followed by a repositioning of the
accretion disk and jets is a plausible interpretation for misaligned radio
structures, even if they are not conspicuously X-shaped \cite{cohen05}.

The radio luminosity of 1045+352 at 1.4\,GHz (Table~\ref{table1}) is high, making this
source one of the most radio-luminous BAL quasars, with a value
similar to that of the first known radio-loud BAL\,QSO with an FR\,II
structure, FIRST\,J101614.3+520916 \cite{gregg00}.
The radio-loudness parameter,
$R^{\ast}$, defined as the {\it K}-corrected ratio of the 5-GHz radio flux
to 2500\AA~optical flux \cite{stocke92} has been calculated for 1045+352
and amounts to $\log(R^{\ast})=4.9$.
For this, a global radio spectral index, $\alpha_{radio}=-0.8$ and an
optical spectral index, $\alpha_{opt}=-1.0$, have been assumed and the SDSS
$g'$ magnitude has been converted to the
Johnson-Morgan-Cousins {\it B} magnitude using the transformation formula
\cite{smith02}.
Corrections have also been made for intrinsic extinction (local to the
quasar) calculated assuming a Milky-Way extinction curve \cite{willott02}.
Even after correction, $\log(R^{\ast})>1$ and amounts to
$\log(R^{\ast})=4.1$, which means that 1045+352 is still radio-loud object.
The two-sided asymmetric structure of 1045+352 can indicate the source
grows up in an asymmetric environment which is very typical among small scale
CSS sources \cite{saikia01}.
However, it was also found \cite{saikia01,
jeyakumar05} that the radio properties of CSS sources are consistent
with the unified scheme in which the axes of the quasars are observed close
to the line of sight. It is then likely that in the case of the BAL
quasar 1045+352 the orientation effect is substantial.

\section{Conclusions}

1045+352 is a very radio-luminous BAL quasar,
whose complex structure is suggestive of restarted activity and may have
resulted even from a minor merger. 
The radio morphology of 1045+352 shows also that the radio jets and BALs can
coexist, at least during some stage of the quasar lifetime. According to
\cite{lipari06}, in radio-loud systems, the jets then remove the clouds
responsible for the generation of BALs, so the extended radio structures
showing BAL features are very rare.

1045+352 is also a CSS object.
CSSs constitute an intermediate class in an evolutionary sequence of radio
sources between smaller GPS objects and very extended Large Symmetric
Objects (LSO). They are young sources with a typical age of $\sim10^{5}$ years
\cite{mur99}, which is an argument in favour of an evolutionary scenario
for BAL quasars, in which the BAL features appear at a very early stage in
the quasar evolution.

\section*{Acknowledgement}
This work was supported by Polish Ministry of Education and Science
under grant 1 P03D 008 30.

\end{document}